\documentclass[12pt,a4paper]{article}
\usepackage{a4wide}
\usepackage{amsfonts}
\begin{document}
{\renewcommand{\thefootnote}{\fnsymbol{footnote}}
\mbox{}\hfill  PITHA -- 99/35\\
\mbox{}\hfill FSUJ-TPI-12/99\\
\mbox{}\hfill ESI 802

\bigskip

\begin{center}
{\LARGE Symplectic Cuts and Projection Quantization}\\
\vspace{1.5em}
Martin Bojowald$^a$\footnote{e-mail address: {\tt
bojowald@physik.rwth-aachen.de}} and Thomas Strobl$^b$\footnote{e-mail
address: {\tt Thomas.Strobl@tpi.uni-jena.de}}
\\\vspace{0.5em}
$^a$Institut f\"ur Theoretische Physik, RWTH Aachen, D--52056 Aachen,
Germany\\\vspace{0.5em} $^b$Institut f\"ur Theoretische Physik,
Universit\"at Jena, D--07743 Jena, Germany\\
\vspace{1.5em}
\end{center}
}

\setcounter{footnote}{0}

\newcommand{\md}{\mathrm{d}}
\newcommand*{\id}{\mathrm{1\mkern-4.3mu I}}
\newcommand*{\R}{{\mathbb R}}
\newcommand*{\N}{{\mathbb N}}
\newcommand*{\Z}{{\mathbb Z}}
\newcommand*{\C}{{\mathbb C}}
\newcommand{\CP}{{\cal{P}}}

\begin{abstract}
  The recently proposed projection quantization, which is a method to
  quantize particular subspaces of systems with known quantum theory, 
  is shown to yield a genuine quantization in several cases. This may
  be infered from exact results established within symplectic cutting.
\end{abstract}

\section{Introduction}

Motivated by studying the phase space $S^1\times\R^+$, which is
defined to be the restriction of $T^*S^1$ to positive momentum, we
recently proposed the projection quantization \cite{project}. The
conditions for its applicability were formulated as:

\begin{enumerate}
\item The phase space ${\cal P}$, which is to be quantized, can be
  characterized as a submanifold of a phase space $\widetilde{\cal P}$
  via restriction by means of inequalities $f_i>0$ for a set of
  functions $\{f_i\}$ on $\widetilde{\cal P}$ with mutually vanishing
  Poisson brackets. Furthermore, for each $i$ the set on which the
  opposite inequality, $f_i<0$, is fulfilled has to be nonempty. For
  simplicity we assume that ${\cal P}$ is connected.
\item A quantum realization of $\widetilde{\cal P}$ is known in which
  the functions $f_i$ may be promoted to self-adjoint, simultaneously
  diagonalizable operators $\hat{f}_i$.
\end{enumerate}

For elementary examples this quantization scheme has been shown to
yield the expected results in Secs.\ II.E and III.B.2 of Ref.\ 
\cite{project}. The aim of the present paper is to show its validity
for a larger class of phase spaces which are specified more precisely
below.

To quantize ${\cal P}$ by means of projection quantization one starts
from the given quantization of $\widetilde{\cal P}$ with the operators
$\hat{f}_i$ acting on the Hilbert space $\widetilde{\cal H}$. These
operators, being required to be self-adjoint and simultaneously
diagonalizable, have mutually commuting spectral families, which can
be used to construct a projector $P$ to the positive part of the
spectra of all the $\hat{f}_i$. To that end, we need simultaneous
diagonalizability of the operators and mere commutativity on a dense
domain would not suffice. The Hilbert space for the quantization of
${\cal P}$ is defined to be the projection ${\cal H}:=P\widetilde{\cal
  H}$. Moreover, the projection can be used to project operators on
$\widetilde{\cal H}$ to operators on ${\cal H}$ as quantized
observables. Note, however, that adjointness properties of those
operators are conserved only under certain conditions. E.g., the
projection of a self-adjoint operator on $\widetilde{\cal H}$ is in
general symmetric on ${\cal H}$, but not necessarily self-adjoint (see
Ref.\ \cite{project} for details).

The condition that the set determined by $f_i<0$ is nonempty is
introduced to exclude systems like $T^*(\R^2\backslash\{0,0\})$. Using
$x$ and $y$ as coordinates of $\R^2$, this phase space can be viewed
as subspace of $T^*\R^2$ subject to the condition $x^2+y^2>0$. The set
$x^2+y^2<0$ is, of course, empty.  Using a standard quantization of
$T^*\R^2$, zero lies in the continuous part of the spectrum of a
quantization of $x^2+y^2$, implying that the projector to its positive
part is the identity.  There would, therefore, be no difference in the
quantum theories of $T^*(\R^2\backslash\{0,0\})$ and $T^*\R^2$.  In
particular, the $\theta$-angle, which for this phase space is of
physical relevance as demonstrated by the Aharonov--Bohm experiment,
cannot be obtained. Such a failure can, however, also occur if the
requirements of projection quantization {\em are\/} fulfilled. For
instance, we can change the above condition to $x^2+y^2>a$ with some
positive number $a$. Then the projector of projection quantization will
be nontrivial, but we will not obtain the $\theta$-angle. This
behavior is generic if the circle action has fixed points.

The above condition on the functions $f_i$ can be reformulated more
precisely as requiring that zero be a regular value of all $f_i$,
i.e.\ that $\md f_i$ is nonzero for each $i$ on the pre-image of zero.
(It is possible to weaken this condition, e.g.\ by demanding that the
moment map of the torus action generated by all the $f_i$ has zero as
a regular value. Below we will, however, impose the conditions $f_i>0$
in steps, which means that each of them is treated as a single
constraint. More generally, one could use multiple cutting \cite{M2}
to deal with the complete torus action.)  The set $f_i^{-1}(0)$ for
each fixed $i$ is then a reducible splitting hypersurface of
$\widetilde{\cal P}$, i.e.\ it is an oriented hypersurface of
codimension one with a free action generated by the Hamiltonian vector
field of $f_i$ and it splits $\widetilde{\cal P}$ in two disjoint open
pieces $\widetilde{\cal P}_+=f_i^{-1}(0,\infty)$ and $\widetilde{\cal
  P}_-=f_i^{-1}(-\infty,0)$ such that its positive normal vectors
point into $\widetilde{\cal P}_+$ and negative normal vectors into
$\widetilde{\cal P}_-$ (see Ref.\ \cite{presymp}).

In case of a single condition $f>0$ which generates a free circle
action on $\widetilde{\cal P}$ we can employ the symplectic cutting
technique\footnote{We are grateful to A.\ Alekseev for bringing this
  method to our attention.} \cite{scuts} to reformulate it as a
constraint $\phi=0$ on an extended phase space $\widetilde{\cal
  P}\times\C$. Here, $\C$ is endowed with the symplectic structure
$\omega_{\C}=-\frac{i}{2}\md z\wedge\md\overline{z}$. If we denote the
circle action on $\widetilde{\cal P}$ by $S^1\colon p\mapsto
e^{it}\cdot p=\exp(tX_f)p$ ($X_f$ being the Hamiltonian vector field
of $f$), we have the free circle action $S^1\colon(p,z)\mapsto
(e^{it}\cdot p,e^{-it}z)$ on $\widetilde{\cal P}\times\C$ with
momentum map $\phi:=f-\frac{1}{2}|z|^2$. The cut space
$\widetilde{\cal P}_{\rm cut}$ is defined as the reduced phase space
$\phi^{-1}(0)/S^1$ subject to the constraint $\phi=0$. It contains
both the reduced phase space $\widetilde{\cal P}_{\rm red}$ of
$\widetilde{\cal P}$ subject to the constraint $f=0$ and the subspace
${\cal P}=\widetilde{\cal P}_+$ we are interested in. Due to the fact
that $\widetilde{\cal P}_{\rm cut}$ is obtained by gluing
$\widetilde{\cal P}_{\rm red}$ into $\widetilde{\cal P}_+$ it has in
general topological properties different from ${\cal P}$. E.g., if
$\widetilde{\cal P}$ is a cylinder, which is rotated by the circle
action, then ${\cal P}$ is a half-cylinder, which is not
simply-connected, whereas $\widetilde{\cal P}_{\rm cut}$ is
simply-connected with the finite boundary compactified to a single
point. Such a topological difference will change the quantum theory,
and therefore we will use an alteration of the standard symplectic
cutting which leads to a phase space not containing the reduced phase
space $\widetilde{\cal P}_{\rm red}$.

Obviously, $\widetilde{\cal P}_{\rm red}$ appears in
$\widetilde{\cal P}_{\rm cut}$ as reduction of $\widetilde{\cal
  P}\times\{0\}\subset\widetilde{\cal P}\times\C$. It can then easily
be seen that the altered symplectic cutting starting from the phase
space $\widetilde{\cal P}\times\C^*$, $\C^*=\C\backslash\{0\}$, leads
to $\widetilde{\cal P}_+={\cal P}$ as symplectic manifolds. We will
use this symplectic cutting when dealing with circle actions in
Section \ref{s:circle}. Translating it into a Dirac quantization will
enable us to prove that for circle actions projection quantization
leads to results equivalent to a quantization starting directly from
${\cal P}$.

The main idea goes as follows: At least under specific conditions, it
has been proven (for further details and references cf.\ Section
\ref{Dirac}) that the Dirac quantization of a constraint $\phi = 0$
generating a circle action yields the same result as the quantization
of the respective reduced phase space (``quantization and reduction
commute''). Starting from $\widetilde{\cal P}\times\C^*$, the
symplectic reduction with respect to $\phi=0$ yields nothing but the
phase space $\CP$ we are interested in. So, in order to prove that
projection quantization yields a genuine quantization of $\CP$, it
suffices to show that it yields a quantum theory equivalent to the one
obtained in a Dirac quantization of $\widetilde{\cal P}\times\C^*$.
This, however, is quite easy to show.

Excision of the origin of $\C$ is necessary also in order to
generalize this construction to actions of the real line with no
closed orbits in Section \ref{s:line}.

In the Discussion we will discuss examples for these methods and
also present an example with a single function $f$ which fulfills the
assumptions of projection quantization but for which the above results
do not apply.  We will finally comment on some possible
generalizations of these considerations.

\section{Projection Quantization with Circle Actions} 
\label{s:circle}

As said above, we will translate symplectic cutting into a Dirac
quantization, i.e.\ we will start by quantizing the phase space
$\widetilde{\cal P}\times\C^*$ followed by imposing the constraint
$\phi=0$ at the quantum level. A necessary ingredient of this
procedure is the quantization of $\C^*$ with its observable
$\frac{1}{2}|z|^2$, which will be presented first in terms of
geometric quantization (the quantization of $\widetilde{\cal P}$ is
assumed to be known). 

\subsection{Geometric Quantization of $\C^*$}

The standard symplectic structure of $\C^*$ is given by
$\omega=-\frac{i}{2}\md z\wedge\md\overline{z}$. Writing $z=q+ip_q$,
we can see that the observable $\frac{1}{2}|z|^2=
\frac{1}{2}(q^2+p_q^2)$, which we have to quantize, is the Hamiltonian
of the harmonic oscillator (with removed origin $q=p_q=0$).
Introducing polar coordinates $z=re^{i\varphi}$ shows that this phase
space is symplectomorphic to $S^1\times\R^+$ with symplectic structure
$\omega=r\md\varphi\wedge\md r=\md\varphi\wedge\md p$, where the
momentum $p=\frac{1}{2}r^2=\frac{1}{2}|z|^2$ is introduced.  The group
theoretical quantization of this phase space has been studied in
detail in Refs.\ \cite{Schramm,project} (see also Ref.\ \cite{Loll};
the phase space also plays an important role in quantum
optics\footnote{We thank H.\ Kastrup for this remark and related
  discussions.} \cite{Lynch}) together with the quantization of the
observable $p$. In fact, this example motivated the definition of
projection quantization.  However, the group theoretical quantization
lead to a quantum ambiguity (which is expected because the phase space
is not simply connected) parameterized by a parameter $k\in\R^+$
(stemming from the positive discrete series of the
$so(2,1)$--representations), whereas projection quantization was seen
to lead more naturally to a parameter $k\in(0,1]$. Therefore, we use
here an independent geometric quantization to decide which domain to
use for the parameter.  Furthermore, we prefer geometric quantization
in this context because our later argumentation will be based
completely on this scheme.

Noting the similarity to the harmonic oscillator we can quantize our
phase space along the lines of this example following Ref.\ 
\cite{Simms} to which we refer for details (see also Refs.\ 
\cite{GuiStern,Woodhouse,geom}; we use Simms' quantization of the
harmonic oscillator and not the more usual quantization using a
holomorphic polarization because it is then straightforward to include
the $\theta$-angle for $\C^*$). Differences to Simms' treatment will
only occur because of a possible $\theta$-angle and when choosing the
metaplectic structure.

In polar coordinates the symplectic form is
$\omega=r\md\varphi\wedge\md r=\md\Theta$ with symplectic potential
$\Theta=-\frac{1}{2}r^2\md\varphi+\hbar\theta\md\varphi$.  Here the
$\theta$-angle appears because the phase space is not simply
connected. As polarization we choose the one generated by the
Hamiltonian vector field $\frac{\partial}{\partial\varphi}$ of the
observable $\frac{1}{2}r^2$. We use the canonical metaplectic
structure associated with this polarization \cite{Woodhouse}. Using
the trivial Hermitean line bundle, wave functions can be written as
$\psi=f\cdot s\otimes\nu$ with a function $f\colon\C^*\to\C$, the unit
section $s$ of the prequantum line bundle, and a constant half-form
$\nu$ satisfying $L_{\frac{\partial}{\partial\varphi}}\nu=0$. This
leads to the polarization condition
\[
  \nabla_{\frac{\partial}{\partial\varphi}}f\cdot s=
  \frac{\partial f}{\partial\varphi}\cdot s+
  \frac{i}{\hbar}f\,\Theta\left(\frac{\partial}{\partial\varphi}\right)s=
  \left(\frac{\partial f}{\partial\varphi}-\frac{i}{2\hbar} 
   r^2f+i\theta f\right)s=0
\]
which has only distributional solutions\footnote{This is to be
  expected for a polarization with compact leaves \cite{Woodhouse}.}
proportional to
\[
  f_n(r,\varphi)=\delta\left(r-\sqrt{2\hbar(n+\theta)}\right)
  e^{in\varphi}\quad,\quad n\in\Z\,.
\]
Because the label $n$ is restricted to satisfy $n+\theta>0$, we can
restrict the parameters to lie in $\theta\in(0,1]$ and $n\in\N_0$ in
order to obtain a family of inequivalent quantizations labeled by the
parameter $\theta$. For each fixed $\theta$ we will denote the Hilbert
space abstractly generated by all $f_n$ as ${\cal H}_{\theta}$.

The observable $p=\frac{1}{2}r^2=\frac{1}{2}|z|^2$ acts on polarized
states just by multiplication
\[
  {\textstyle\frac{1}{2}}r^2
  f_n(r,\varphi)=\hbar(n+\theta)f_n(r,\varphi)
\]
with spectrum $\{\hbar(n+\theta):n\in\N_0\}$. Comparing with the
spectrum for $p$ obtained within the methods of Refs.\ 
\cite{Schramm,project}, we see that $\theta\in(0,1]$ leads to results
equivalent to projection quantization, whereas the group theoretical
quantization leads to a larger class of inequivalent quantum
realizations.

We complete this discussion with a remark on the metaplectic structure
(see Ref.\ \cite{Woodhouse} for details). Due to $H^1(\C^*,\Z_2)=\Z_2$
there are two inequivalent metaplectic structures on the phase space.
The structure different from the one used above can be obtained by
restricting the canonical metaplectic structure of $T^*\R$ to the
subspace $\C^*=T^*\R\backslash\{(0,0)\}$. As is well known from the
harmonic oscillator, this leads to a metaplectic correction in the
spectrum of the Hamiltonian providing the zero point energy.  The
spectrum is then $\{\hbar(n+\frac{1}{2})\}$.  In the above
quantization we chose the canonical metaplectic structure associated
with the polarization $\frac{\partial}{\partial\varphi}$, which
appears to be more natural when interpreting the phase space as
$S^1\times\R^+$ with the observable $p$. In this case there is no
metaplectic correction.

\subsection{Dirac Quantization and Symplectic Cuts}
\label{Dirac}

In the preceding subsection we have shown that geometrical and
projection quantization lead to equivalent results for the phase
space $\C^*$; in particular, both schemes yield the same domain for
the $\theta$-angle. We will now extend this result to a larger class
of phase spaces by using symplectic cutting. As in the classical
framework (manipulating symplectic manifolds) the phase space $\C^*$
and its quantization play an important role.

Projection quantization is devised to the quantization of phase spaces
${\cal P}$ which are subspaces of a larger phase space
$\widetilde{\cal P}$ given by suitable conditions $f_i>0$. It can be
helpful for phase spaces ${\cal P}$ which are complicated to quantize
explicitly, but which are embedded into a phase space $\widetilde{\cal
  P}$ with a well understood quantization (e.g., $\widetilde{\cal P}$
could be a cotangent bundle; the method is, however, not restricted to
this case). One then starts from the known quantization of
$\widetilde{\cal P}$ and projects to a subspace of the quantum Hilbert
space to obtain the Hilbert space of ${\cal P}$. The basic idea to
prove that this will, under certain conditions, lead to the correct
result is to use Dirac quantization of symplectic cutting. As main
ingredient into this proof we use theorems which state the commutation
of reduction and quantization which means that a quantization of the
reduced phase space of a constrained system is equivalent to the space
annihilated by all constraint operators represented on a Hilbert space
quantizing the unreduced phase space of the system.  Such theorems go
back to a conjecture of Guillemin and Sternberg \cite{GS} in case of
compact K\"ahler manifolds, and have recently been proved and extended
using symplectic cutting (see, e.g.\ Refs.\ \cite{DGMW,M2,presymp}, in
Ref.\ \cite{M1} the arguments are generalized to non--K\"ahler
manifolds).  Independently, this has been investigated in Refs.\ 
\cite{Stillerman,Gotay} for not necessarily compact phase spaces which
are physically more interesting. The main condition for commutation of
reduction and quantization is that the action of the gauge group $G$
generated by the constraints preserves the structure used to quantize
$\widetilde{\cal P}$, namely the polarization and the metaplectic
structure. Then one can project the quantization structure to the
reduced phase space and establish the commutation of reduction and
quantization in a controled way. In the following we restrict our
considerations to the class of phase spaces where the conditions for a
commutation of reduction and quantization are fulfilled. The only
requirement on the symplectic geometry of a particular phase space is
that it has to permit a polarization and a metaplectic structure
compatible with the constraints in the above sense.

In the present section we first treat phase spaces $\widetilde{\cal
  P}$ with a single constraint $f$ which generates a free circle
action on $\widetilde{\cal P}$ with momentum map $f$ such that zero is
a regular value of $f$, and will later generalize to torus actions. 
By assumption, furthermore, the quantum theory
of $\widetilde{\cal P}$ and its Hilbert space $\widetilde{\cal H}$ are
known. (If the quantization is not unique, we can use any of the
inequivalent quantum realizations of $\widetilde{\cal P}$.) In the
preceding section we derived the Hilbert spaces ${\cal H}_{\theta}$
for the quantum theory of $\C^*$, which we use to construct the
Hilbert space $\widetilde{\cal H}\otimes{\cal H}_{\theta}$ of
$\widetilde{\cal P}\times\C^*$ for arbitrary $\theta\in(0,1]$.
Together with the Hilbert space $\widetilde{\cal H}$ we assume to know
a self-adjoint quantization $\hat{f}$ on $\widetilde{\cal H}$ of the
function $f$. Combining this with the quantization of
$\frac{1}{2}|z|^2$ acting on ${\cal H}_{\theta}$ we obtain the
quantized constraint
\[
  \hat{\phi}=\hat{f}\otimes\id-
  \id\otimes{\textstyle\frac{1}{2}}\widehat{|z|}^2
\]
acting on $\widetilde{\cal H}\otimes{\cal H}_{\theta}$, which imposes
symplectic cutting at the quantum level.

Provided that quantization and reduction commute, 
the quantization of ${\cal P}=\widetilde{\cal P}_+$ is given by the
kernel of $\hat{\phi}$. We know the spectrum of
$\frac{1}{2}\widehat{|z|}^2$ from the preceding subsection, where we
showed that it is discrete.  Similarly, the spectrum of $\hat{f}$ is
discrete: $f$ generates a Hamiltonian circle action on
$\widetilde{\cal P}$ which, provided that $f$ is quantizable in the
sense of geometric quantization, implies that $\hat{f}$ generates a
unitary (possibly projective) $S^1$-representation on $\widetilde{\cal
  H}$. This representation splits into sectors $\widetilde{\cal
  H}_{\theta_j}$ on which $e^{2\pi i \varphi}\in S^1$ has eigenvalues
of the form $e^{2\pi i (n+\theta_j) \varphi}$, $\theta_j\in(0,1]$,
with integer values of $n$.  In other words, in each of these sectors,
$\hat{f}$ has as spectrum a subset of $\{n+\theta_j:n\in\Z\}$.

Let us first assume that there is only one $\theta_j=\theta'$, i.e.\ 
$\widetilde{\cal H}=\widetilde{\cal H}_{\theta'}$ (this is always the
case if the algebra of observables is represented irreducibly on
$\widetilde{\cal H}$, because the elements of the fundamental group of
$\widetilde{\cal P}$ are required to commute with all observables; for
a discussion cf, e.g. Ref.\ \cite{Giulini}). Then for
$\theta\not=\theta'$ the kernel of $\hat{\phi}$ is trivial, whereas
for $\theta=\theta'$ the constraint $\hat{\phi}=0$ acting on
$\psi\otimes f_n\in \widetilde{\cal H}\otimes{\cal H}_{\theta}$ takes
the form $\hat{f}\psi=(n+\theta)\psi$ with $n\in\N_0$. This constraint
projects precisely to those states of $\widetilde{\cal H}$ which are
eigenstates of $\hat{f}$ with {\em positive\/} eigenvalues, leading
exactly to the result of projection quantization.

If there are different values of $\theta_j$ in $\widetilde{\cal H}$
(the algebra of observables is then represented reducibly),
we have to match all the $\theta$-sectors separately by choosing an
appropriate direct sum of ${\cal H}_{\theta}$ as quantization of
$\C^*$.  

These considerations can easily be extended to the case of more than
one commuting conditions $f_j>0$. We can reduce them one after
another, not running into problems because their actions commute and
therefore project to the cut spaces.  If, moreover, the commutation
assumption on quantization and reduction is fulfilled for the
constraints $f_j=0$ on $\widetilde{\cal P}$, it also holds for
$\phi_j=0$ on $\widetilde{\cal P}\times\C^*$. This is a consequence of
the above construction, where we always took direct products of
quantization data. They are conserved by the circle actions generated
by $\phi_j$ provided the data on $\widetilde{\cal P}$ are conserved by
the actions generated by $f_j$ (it is immediate to see that this is
also fulfilled for the constraint $\frac{1}{2}|z|^2$ on $\C^*$ in the
above quantization).

We thus may conclude

\smallskip

{\bf Theorem 1 (Projection Quantization with Circle Actions):} {\it
  Let the functions $f_j$ on a phase space $\widetilde{\cal P}$, with
  zero being a regular value for each of them, generate mutually
  commuting circle actions.  Assume further that their quantizations
  $\hat{f}_j$ on the Hilbert space $\widetilde{\cal H}$ generate
  mutually commuting unitary actions and that reduction commutes with
  quantization for each of them.

  Then projection quantization applied to $\widetilde{\cal P}$ with the
  conditions $f_j>0$ yields a quantization of ${\cal
    P}=\widetilde{\cal P}_+$.
}

\smallskip

Let us emphasize that our general assumption of commutation of
quantization and reduction has to be imposed only for the constraints
$f_j$. As such, this is a statement about the {\em reduced\/} phase
space $\widetilde{\cal P}_{\rm red}$ (characterized by $f_j=0$)
and its quantization.  Its validity can in many cases be seen by
employing the results collected in the references mentioned above or
by checking it by hand in specific cases.  As the construction
shows, we can take then for granted that reduction commutes with
quantization also for each constraint $\phi_j$ on $\widetilde{\cal
  P}\times\C^*$ associated with $f_j$. This allowed us to prove the
desired result without requiring a condition (like the commutation
assumption) for the constraints $\phi_j$ or the phase space
$\widetilde{\cal P}_+$.

As already discussed in the Introduction, the quantization obtained
using the present formulation of projection quantization is not always
the most general one.  In particular, if the fundamental group of
$\widetilde{\cal P}_+$ does not coincide with the one of
$\widetilde{\cal P}$ (there can be contractible loops in
$\widetilde{\cal P}$ which become noncontractible after imposing
$f_i>0$), some of the $\theta$--angles necessary for a general
quantization of $\widetilde{\cal P}_+$ may be missed. At least in some
cases an adaption of the quantization scheme may accomodate for these
additional $\theta$--angles; we intend to come back to this issue
elsewhere.  If, on the other hand, any two nonhomotopic loops in
$\widetilde{\cal P}_+$ are nonhomotopic also as loops in
$\widetilde{\cal P}$, the $\theta$--angles in the most general
quantization of $\widetilde{\cal P}$ are sufficient to yield the most
general quantum theory of $\widetilde{\cal P}_+$.  This is e.g.\ the
case in the paradigmatic example of a cylinder cut to a
half--cylinder.

\subsection{Observables} 

Dirac quantization also contains a prescription to obtain observables
on the physical Hilbert space ${\cal H}_1$, on which the constraints
are solved, from those on the original Hilbert space ${\cal
  H}_0$. An operator ${\cal O}_0$ on ${\cal H}_0$ which is an
observable in the strict sense, i.e.\ which commutes with the quantum
constraint operator $\hat \phi$ and thus with the projector $P$ to
${\cal H}_1$, projects just to the same operator restricted to ${\cal
  H}_1$.

More generally, one can project any operator ${\cal O}_0$ on ${\cal
  H}_0$ to an operator $P{\cal O}_0P$. The new operator annihilates
the orthogonal complement of ${\cal H}_1$ and can, therefore, be
reduced to an operator on this subspace. If we denote the inclusion of
the subspace ${\cal H}_1$ into ${\cal H}_0$ as $\iota_0\colon{\cal
  H}_1\hookrightarrow{\cal H}_0$ and the projection from ${\cal H}_0$
to ${\cal H}_1$ as $\pi_0\colon{\cal H}_0\to{\cal H}_1$, we can write
the final operator on ${\cal H}_1$ as ${\cal O}_1=\pi_0\circ{\cal
  O}_0\circ\iota_0\colon{\cal H}_1\to{\cal H}_1$.

In our case we have ${\cal H}_0=\widetilde{\cal H}\otimes{\cal
  H}_{\theta}$ and $P$ is the projector to the kernel ${\cal H}$ of
the constraint $\hat{\phi}$. The quantum theory on $\widetilde{\cal H}$
is assumed to be known, and therefore we have observables ${\cal O}$
acting on this Hilbert space. They can be extended trivially to
operators ${\cal O}\otimes\id$ on ${\cal H}_0$ and, using the
procedure described above, projected to operators on ${\cal H}$.

This leads exactly to the definition of observables which has been
given in Ref.\ \cite{project} in the framework of projection
quantization.  Thus, Dirac quantization of symplectic cutting leads,
under the conditions stated in the Theorem, to the same observables as
projection quantization.

Using the projection of operators, we can associate an operator ${\cal
  O}$ on the Hilbert space ${\cal H}$, which quantizes ${\cal P}$, to
each operator $\widetilde{\cal O}$ on $\widetilde{\cal H}$. If
$\widetilde{\cal O}$ is the quantization of a phase space function on
$\widetilde{\cal P}$, ${\cal O}$ can be regarded as quantization of
the same phase space function restricted to ${\cal
  P}\subset\widetilde{\cal P}$. However, even if $\widetilde{\cal O}$
is selfadjoint (or unitary), ${\cal O}$ will be selfadjoint (unitary)
in general only if $\widetilde{\cal O}$ is an observable in the strict
sense of Dirac quantization, i.e.\ if it commutes with the projector
to the physical subspace. Otherwise, ${\cal O}$ is in general only
hermitean (isometric). A more detailed discussion has been given in
Ref.\ \cite{project}.

\section{Projection Quantization with Line Actions}
\label{s:line}

We now indicate how the results of the preceding section can be
generalized to the case that the orbits of the action generated by $f$
do not close but are homeomorphic to $\R$. For technical reasons, we
restrict our considerations to conditions $f>0$ where $f$ is chosen to
be a coordinate on $\widetilde{{\cal P}}$. First we have to adapt the
symplectic cutting to line actions.

We replace $\C^*$ used before by its universal covering space
$\widetilde{\C^*}$ parameterized by $(r,x)\in\R^+\times\R$ with
covering map $(r,x)\mapsto re^{ix}$. The symplectic form is
$\omega=r\md x\wedge\md r$. The rest of symplectic cutting is as
before with $\phi:=f-\frac{1}{2}r^2$ generating a free action on
$\widetilde{\cal P}\times\widetilde{\C^*}$. The subset ${\cal
  P}=\widetilde{\cal P}|_{f>0}$ is symplectomorphic to
$\phi^{-1}(0)/\R$.

Now we can proceed similar to the case of a circle action by commuting
quantization and reduction.  There are, however, two differences:
First, there is no $\theta$-angle and, second, the spectrum of
$\frac{1}{2}r^2$ is continuous (we can use a quantization similar to
that of $\C^*$, but without the restriction $n\in\Z$). Therefore, zero
will lie in the continuous part of the spectrum of the constraint
$\hat{\phi}$, which leads to technical difficulties when projecting to
its kernel.  In the following we will use group
averaging \cite{ALMMT,Marolf} and assume $f$ to be chosen as coordinate
(otherwise the following calculations have to be adapted
appropriately). In particular, the polarization chosen to quantize
$\widetilde{\cal P}$ contains the Hamiltonian vector field $X_f$ of
$f$ and the symplectic potential $\Theta_{\widetilde{\cal P}}$ on
$\widetilde{\cal P}$ is adapted to $X_f$, i.e.\ 
$\Theta_{\widetilde{\cal P}}(X_f)=0$.  In $\widetilde{\C^*}$ we choose
the polarization generated by $\frac{\partial}{\partial x}$ and
symplectic potential $\Theta=rx\md r=x\md p$, $p=\frac{1}{2}r^2$.
Quantum states of $\widetilde{\cal P}\times\widetilde{\C^*}$ can then
be represented as $\psi(f,y)\chi(p)$, where $\psi$ is a quantum state
of $\widetilde{\cal P}$ depending on $f$ and other continuous or
discrete labels collected in $y$.

The constraint $\hat{\phi}$ generates the unitary $\R$-action
\[
  \psi(f,y)\chi(p)\mapsto e^{it(f-p)}\psi(f,y)\chi(p).
\]
For group averaging we use test states which are smooth and of compact
support. The rigging map $\eta$ is then determined by
\begin{eqnarray*}
  \eta(\psi_1\chi_1)[\psi_2\chi_2] & = & \mu_y\left(\int_{\R}\md
    f\int_{\R^+}\md p\, \nu(y,f)\int_{\R}\md t\,
    e^{it(f-p)}\psi_1(f,y)\chi_1(p)
    \overline{\psi_2(f,y)\chi_2(p)}\right)\\
 & = & \mu_y\left(\int_{\R^+}\md
   p\,\nu(p,y)\psi_1(p,y)\chi_1(p)\overline{\psi_2(p,y)\chi_2(p)}\right)
\end{eqnarray*}
factoring without restriction the measure for polarized states of
$\widetilde{\cal P}$ into $\mu_y\int_{\R}\md f\, \nu(f,y)$. This
calculation demonstrates that the image of the rigging map coincides
with the spectral projection to the positive part of the spectrum of
$f$ (the coordinate $f$ is replaced by the positive coordinate $p$).
Assuming commutation of quantization and reduction, we see that also
for line actions projection quantization yields a quantization of
${\cal P}=\widetilde{\cal P}_+$ where a coordinate is constrained to
be positive:

\smallskip

{\bf Theorem 2 (Projection Quantization with Line Actions):} {\it Let
  $f_j$ be coordinates generating line actions on a phase space
  $\widetilde{\cal P}$ which is equipped with a polarization
  containing the vector fields generated by the $f_j$ and a symplectic
  potential adapted to the polarization. Assume further that reduction
  commutes with quantization for each of the $f_j$ regarded as
  constraints.

  Then projection quantization applied to $\widetilde{\cal P}$ with the
  conditions $f_j>0$ yields a quantization of ${\cal
    P}=\widetilde{\cal P}_+$.
}

\section{Discussion}

The result of this paper is a proof that projection quantization
proposed in Ref.\ \cite{project} leads to a correct quantization in
particular cases. The results apply in case of circle and line actions
generated by functions $f_i$ on $\widetilde \CP$, with the main
assumption that reduction and quantization with respect to $f_i=0$
commute (as has been proven for a fairly general class of systems,
cf.\ the citations above).

The systems covered by our requirements provide further examples
supplementary to those of Ref.\ \cite{project} in which projection
quantization leads to the correct results. These include e.g.\ the
following class of simple systems: $T^*(S^1)^n$ or $T^*\R^n$ with some
of the momenta or coordinates (for $T^*\R^n$) constrained to be
positive. More generally, arbitrary linear combinations of the
coordinates of $T^*\R^n$ or linear combinations with integer
coefficients of the momenta of $T^*(S^1)^n$ can be constrained to be
positive.  But certainly many more examples are covered by the present
considerations, e.g.\ in cases where a globally defined coordinate on
a cotangent bundle is constrained to be positive and one uses the
vertical polarization. For instance, the phase space $T^*GL^+(n,\R)$,
which is the cotangent bundle on the manifold of $n\times n$-matrices
of positive determinant, can be treated along these lines if one
embeds it into the cotangent bundle $T^*M(n,\R)\cong T^*\R^{n^2}$ on
the manifold of all $n\times n$-matrices via the restriction $f:=\det
A>0$ for $A\in M(n,\R)$. One then directly obtains a quantization of
$T^*GL^+(n,\R)$ on the space of square integrable functions on
$GL^+(n,\R)\subset \R^{n^2}$. The fundamental operators are projected
to multiplication and derivative operators the latter of which are no
longer self-adjoint (analogous to the phase space $T^*\R^+\cong
T^*GL^+(1,\R)$).  This system has already been dealt with in a group
theoretical quantization \cite{Isham} leading to the same Hilbert
space besides a large family of ``degenerate'' quantizations. This is
similar to the phase space $S^1\times\R^+$ where group theoretical
quantization yields a larger class of quantizations some of which have
to be regarded as unphysical \cite{project}.  It is, however, not
always possible to select the physical representations intrinsically
from properties of the system, e.g.\ by demanding positive spectra of
suitable operators, and here a comparison with other quantization
schemes, as projection quantization, can help.  Moreover, also in this
case the projection quantization is much simpler to apply than the
group theoretical one; and, according to the results of the present
paper, the application of the projection approach is fully legitimate
here: it yields a quantization equivalent to a genuine geometric
quantization of the restricted space.

Another class of examples where the proofs of this paper can be used
is provided, e.g., by phase spaces which are compact Kaehler
manifolds. For additional conditions in this case we refer to the
literature \cite{M2,presymp,GS,DGMW}.

We now briefly describe a system which is not covered by the 
results of the present article, but can nevertheless be dealt with
using projection quantization.  Let the phase space ${\cal P}$ be the
subspace of the product $\widetilde{\cal P}=T^*S^1\times T^*S^1$
subject to the condition $f:=p_1-p_2^2>0$ in terms of the usual
coordinates $(\varphi_1,p_1)$ and $(\varphi_2,p_2)$ of the two
cylinders. The action on $\widetilde{\cal P}$ generated by $f$ has
orbits winding around the torus $S^1\times S^1$ parameterized by
$(\varphi_1,\varphi_2)$ and they are closed if and only if $p_2$ is
rational. Otherwise they are homeomorphic to $\R$. We have thus
neither of the cases of pure circle or line actions dealt with in the
preceding sections.

Nevertheless, the assumptions of projection quantization as
recapitulated in the Introduction are fulfilled and the method may be
applied for a quantization: Each of the two cylinders in
$\widetilde{\cal P}$ can be quantized in the usual way leading to
quantum states of the form $\psi_{n_1,n_2}(\varphi_1,\varphi_2)=
e^{in_1\varphi_1}e^{in_2\varphi_2}$ (assuming for simplicity vanishing
$\theta$-angles). Projection quantization then selects those states
which have positive eigenvalues for $\hat{f}$:
$\hat{f}\psi_{n_1,n_2}=(n_1-n_2^2)\psi_{n_1,n_2}$ resulting in
the condition $n_2^2<n_1$ for quantum states of ${\cal
  P}=\widetilde{\cal P}_+$.

Of course, it would be of interest to extend the proofs presented here
for circle and line actions to a more general class of actions,
including the above system.  Already this relatively simple example of
a phase space $\CP$ is complicated to quantize by standard methods,
while projection quantization is almost trivial to apply.

These remarks also apply for imposing the condition of a nondegenerate
metric of fixed signature in quantum gravity (see e.g.\ Ref.\ 
\cite{Loll2}).  In this context one in addition has to take into
account that one is dealing with a constrained system. The constraint
$\phi=0$, imposing the condition $f>0$, then arises in addition to the
usual constraints of the gravity system. Consistency leads to
compatibility conditions between the original constraints and $\phi$
(see also the remarks in Ref.\ \cite{project}). As shown independently
in Ref.\ \cite{Klauder}, methods similar to those of projection
quantization can also be used to solve constraints by ``thickening'' a
constraint surface given, e.g., by $f=0$ to a set given by
$-\epsilon<f<\epsilon$ (which corresponds to a twofold application of
projection quantization).

As the preceding example demonstrates, projection quantization can be
helpful for the quantization of physically interesting systems, even
if the above theorems do not apply. Quantization schemes are justified
usually by showing that they yield the expected results for standard
test models. Some elementary systems have been dealt with on these
grounds already in Ref.\ \cite{project}. A large class of further
systems is covered implicitly by the proofs presented in this article.
So we can trust the method also for more complicated systems. Still,
further tests and possibly adaptions of the method are of interest.

The main advantage of projection quantization is that it is extremely
simple {\em if\/} it applies, i.e.\ if its assumptions are fulfilled.
Its applicability, however, is smaller than that of a typical quantization 
procedure: The phase space of interest has
to be embedded into a larger phase space with known quantization,
where the kind of embedding is also restricted by the requirements on
the functions $f_i$. 

In this context, extensions of the symplectic cutting method to
actions of nonabelian groups \cite{M2} can be of interest. This may
lead to a generalization of projection quantization to noncommuting
conditions $f_j$, which would allow a more general class of embeddings
of the phase space.

\section*{Acknowledgements}

We thank A.\ Alekseev and H.\ Kastrup for discussions and J.\ Klauder
for drawing our attention to Ref.\ \cite{Klauder}, in which a related
method has been proposed prior to us (cf.\ also the remark in our
conclusions).  T.S.\ is grateful to the Erwin Schr\"odinger Institute
in Vienna for hospitality in the period when this work was begun and
M.\ B.\ to A.\ Wipf and the TPI in Jena for hospitality during the
time when the work was completed.

\end{document}